\documentclass[aps, prl, preprint]{revtex4-1} 
\usepackage[english]{babel}
\usepackage{amsmath,amsthm}
\usepackage{amsfonts}
\usepackage{xspace}
\usepackage{bm}
\usepackage[squaren, thinqspace]{SIunits}
\usepackage{booktabs}
\usepackage{graphicx}
\usepackage{color}
\theoremstyle{definition}

\theoremstyle{remark}
\begin{document}
\renewcommand{\degree}{\ensuremath{^\circ}\xspace}
\renewcommand{\Re}{\ensuremath{\mathrm{Re}} \xspace}
\renewcommand{\Im}{\ensuremath{\mathrm{Im}} \xspace}
\newcommand{\Hy}{\ensuremath{\bm{H} || \bm{y}}\xspace}
\newcommand{\Hx}{\ensuremath{\bm{H} || \bm{x}}\xspace}
\newcommand{\y}{\ensuremath{\bm{y}}\xspace}
\newcommand{\x}{\ensuremath{\bm{x}}\xspace}
\newcommand{\z}{\ensuremath{\bm{z}}\xspace}
\newcommand{\Smag}{\ensuremath{\left|S_{21}\right|}\xspace}
\newcommand{\DSmag}{\ensuremath{\Delta\left|S_{21}\right|}\xspace}
\newcommand{\ii}{\ensuremath{\mathrm{i}}\xspace}
\newcommand{\LNO}{LiNbO$_3$\xspace}
\newcommand{\Pidt}{\ensuremath{P_\mathrm{IDT}}\xspace}
\newcommand{\Vdc}{\ensuremath{V_\mathrm{DC}}\xspace}
\newcommand{\DPidt}{\ensuremath{\Delta P_\mathrm{IDT}}\xspace}
\newcommand{\Pemw}{\ensuremath{P_\mathrm{EMW}}\xspace}
\newcommand{\Psaw}{\ensuremath{P_\mathrm{SAW}}\xspace}
\newcommand{\Vmsp}{\ensuremath{V_\mathrm{MSP}}\xspace}
\newcommand{\Vmr}{\ensuremath{V_\mathrm{MR}}\xspace}
\newcommand{\Vsp}{\ensuremath{V_\mathrm{SP}}\xspace}
\newcommand{\Vsse}{\ensuremath{V_\mathrm{SSE}}\xspace}
\newcommand{\bsp}{\ensuremath{b_\mathrm{SP}}\xspace}
\newcommand{\Esp}{\ensuremath{E_\mathrm{SP}}\xspace}
\newcommand{\DVdc}{\ensuremath{\Delta V_\mathrm{DC}}\xspace}
\newcommand{\Vind}{\ensuremath{V_\mathrm{ind}}\xspace}
\newcommand{\DPsaw}{\ensuremath{\Delta P_\mathrm{SAW}}\xspace}
\newcommand{\DPemw}{\ensuremath{\Delta P_\mathrm{EMW}}\xspace}
\newcommand{\DVmsp}{\ensuremath{\Delta V_\mathrm{MSP}}\xspace}
\newcommand{\DVmr}{\ensuremath{\Delta V_\mathrm{MR}}\xspace}
\newcommand{\gSpinMix}{\ensuremath{g_{\uparrow\!\downarrow}}\xspace}
\newcommand{\VANE}{\ensuremath{V_\mathrm{ANE}}\xspace}
\newcommand{\VISHE}{\ensuremath{V_\mathrm{ISHE}}\xspace}
\newcommand{\Js}{\ensuremath{\bm{J}_\mathrm{s}}\xspace}
\newcommand{\mus}{\ensuremath{\bm{\mu}_\mathrm{s}}\xspace}
\newcommand{\Jc}{\ensuremath{\bm{J}_\mathrm{c}}\xspace}
\newcommand{\Jssp}{\ensuremath{J_\mathrm{s}^\mathrm{SP}}\xspace}
\newcommand{\Jssmr}{\ensuremath{J_\mathrm{s}^\mathrm{SMR}}\xspace}
\newcommand{\Jssse}{\ensuremath{J_\mathrm{s}^\mathrm{SSE}}\xspace}
\newcommand{\alphaSH}{\ensuremath{\alpha_{\mathrm{SH}}}\xspace}
\newcommand{\lambdaSD}{\ensuremath{\lambda_{\mathrm{SD}}}\xspace}
\newcommand{\tN}{\ensuremath{t_{\mathrm{N}}}\xspace}
\newcommand{\sigmaN}{\ensuremath{\sigma_{\mathrm{N}}}\xspace}
\newcommand{\tF}{\ensuremath{t_{\mathrm{F}}}\xspace}
\newcommand{\sigmaF}{\ensuremath{\sigma_{\mathrm{F}}}\xspace}
\newcommand{\J}{\ensuremath{\mathrm{J}}\xspace}

\title{Comment on "Detection of Microwave Spin Pumping \\Using the Inverse Spin Hall Effect"}%

\author{Mathias Weiler}
\email{mathias.weiler@nist.gov}
\author{Hans T. Nembach}
\author{Justin M. Shaw}
\author{Thomas J. Silva}
\affiliation{National Institute of Standards and Technology, Boulder, CO, 80305}

\date{\today~~Contribution of NIST, not subject to copyright}%
% ----------------------------------------------------------------

\maketitle
% ----------------------------------------------------------------
In a recent Letter~\cite{Hahn:2013}, Hahn \textit{et al.} reported on the detection of an ac voltage in an yttrium iron garnet (YIG)/platinum (Pt) bilayer under the condition of parametrically excited resonance. The authors observe an ac voltage at the frequency of the magnetization precession which is half the frequency of the microwave excitation. As argued by the authors, this parametric excitation scheme allows them to exclude direct microwave crosstalk as the origin of the ac voltage. However, the authors do not address the possibility of an inductive origin for the observed ac voltage $V_\mathrm{ISHE}(\mathrm{ac})$. In ferromagnetic resonance, an inductive signal at the magnetization precession frequency arises as a consequence of Faraday's law and is routinely exploited in, e.g., stripline or vector-network-analyzer ferromagnetic-resonance experiments~\cite{Kalarickal:2006}. Such an inductive signal stems from the threading of the time-varying magnetic flux originating in the YIG film around the Pt stripe. 

%We note that the experimental configuration utilized in Ref.~\cite{Hahn:2013}, whereby voltage leads are used to couple microwaves out of the Pt stripe, is not conducive to quantitative analysis, given the extremely poor nature of the impedance-matching expected for the described geometry. In particular, the dc resistance of the Pt stripe has little bearing on the impedance mismatch: the capacitance per unit length for the Pt stripe, in combination with the voltage leads, is certainly far in excess of that sufficient to achieve \unit{50}{\ohm} at microwave frequencies.

%Inductive ac voltages arise whenever a time-varying magnetization causes a flux change in an inductive loop and are the cause of the signal observed in Fig.~2(a) in Ref.~\cite{Hahn:2013}. 

According to Faraday's law, time-varying flux $\Phi$ causes an inductive voltage $\Vind=-d\Phi/dt$. From reciprocity, the net flux $\Phi$ generated by a magnetic solid of volume $V$ and magnetization $\bm{M}$ that threads an electrical circuit is given by
\begin{equation}\label{eq:phi}
\Phi=\mu_0 \int_V \bm{h}(\bm{r})\cdot \bm{M}(\bm{r}) dr^3, 
\end{equation}
where $\mu_0$ is the vacuum permeability, $\bm{h}(\bm{r})\equiv\bm{H}(\bm{r},I)/I$ and $\bm{H}(\bm{r},I)$ is the Oersted field generated in the magnetic solid when a current $I$ flows in the electrical circuit. As detailed in Ref.~\cite{Silva:1999}, we approximate $\bm{H}$ with the Karlqvist equations for an Oersted field in close proximity of a uniform current sheet~\cite{Mallinson:1993}. Using the coordinate system from Ref.~\cite{Hahn:2013}, \Vind along $z$ in the ac inverse spin Hall effect (iSHE) measurement configuration as shown in Fig.~4(b) in Ref.~\cite{Hahn:2013} then follows as 
\begin{equation}\label{eq:Vind}
\Vind(t)=-\frac{\mu_0 L d_\mathrm{YIG}}{2}\frac{d M_\mathrm{x}(t)}{dt}\frac{Z_0}{Z_0+\frac{1}{2}R_\mathrm{DC}}\:, 
\end{equation}
where $L$ is the length of the sample exposed to the rf magnetic field, $d_\mathrm{YIG}$ is the YIG thickness and $M_\mathrm{x}$ is the dynamic in-plane component of the magnetization. The last term in Eq.~\eqref{eq:Vind} accounts for losses due to the non-zero dc resistance $R_\mathrm{DC}$ of the Pt film~\cite{Silva:1999}. $Z_0$ is the impedance of the measurement circuit. $M_\mathrm{x}$ is  orthogonal to the in-plane magnetization equilibrium orientation parallel to the $z$-axis.  Note that $M_\mathrm{x}$ is responsible for both inductive and potential ac iSHE signals in the geometry used in Ref.~\cite{Hahn:2013}. Both ac voltages are at the same frequency and are generated along the $z$ direction under the same resonance condition. Thus, inductive and ac iSHE signals have identical experimental signatures. The nature of the excitation (direct or parametric) is not relevant for the observation of inductive signals according to Eq.~\eqref{eq:Vind}.

To estimate the expected inductive signal, we use the parameters given  in Ref.~\cite{Hahn:2013} for the $\phi_\mathrm{H}=10\degree$ configuration: $M_\mathrm{s}=\unit{139}{\kilo\ampere\per\meter}$, $L=\unit{500}{\micro\meter}$, $d_\mathrm{YIG}=\unit{200}{\nano\meter}$, $Z_0=\unit{50}{\ohm}$, $R_\mathrm{DC}=\unit{129}{\ohm}$, $\omega/(2\pi)=f_\mathrm{p}/2=\unit{0.546}{\giga\hertz}$ and precession cone angle $\Theta=5.1\degree$. Assuming circular precession for simplicity, we find $M_\mathrm{x}(t)= M_\mathrm{s} \sin\Theta \cos\Theta \cos (\omega t)$. Eq.~\eqref{eq:Vind} thus evaluates to an ac voltage with peak magnitude $\Vind^0 \approx \unit{1}{\milli\volt}$. 

This voltage is two orders of magnitude larger than the voltage $V_\mathrm{ISHE}(\mathrm{ac})=\unit{6.2}{\micro\volt}$ observed in~\cite{Hahn:2013} for $\phi_\mathrm{H}=10\degree$. We note that the experimental configuration utilized in Ref.~\cite{Hahn:2013}, whereby voltage leads are used to couple microwaves out of the Pt stripe, is not conducive to quantitative analysis, given the extremely poor nature of the impedance-matching expected for the described geometry. While poor impedance-matching could thus explain why the authors of Ref.~\cite{Hahn:2013} measure only such a small voltage, it appears that an inductive signal should overwhelm any possible ac iSHE contributions for their sample. We note that the authors of Ref.~\cite{Hahn:2013} repeated their experiment using a YIG/aluminum (Al) bilayer that should show no ac iSHE signal. While no signal was detected, in accordance with the notion that such experiments are sensitive only to the ac iSHE, such a control test cannot be construed as proof that an inductive signal is absent: any number of experimental artifacts stemming from the poor impedance-matching associated with the experimental geometry, and any reproducibility issues that can arise from such poor microwave coupling or differences in the parametric thresholds among samples, may have resulted in this null result.

It is our conclusion that conclusive evidence for detection of ac spin pumping via the iSHE requires substantial additional support, including determination of the parametric excitation threshold as a function of dc magnetic field, an analysis that includes the possibility of spin-wave generation, a more extensive determination of the impedance mismatch, and a theoretical analysis that considers the magnitude of the inductive signal.

\end{document}